# Rotational state-changing cold collisions of hydroxyl ions with helium

Daniel Hauser, Seunghyun Lee, Fabio Carelli, Steffen Spieler, Olga Lakhmanskaya, Eric S. Endres, Sunil S. Kumar, Franco Gianturco, Roland Wester*

*Institut für Ionenphysik und Angewandte Physik, Universität Innsbruck, Technikerstraße 25, 6020 Innsbruck, Austria*

**Cold molecules are important for many applications[1], from fundamental precision measurements[2], quantum information processing[3], quantum-controlled chemistry[4], to understanding the cold interstellar medium[5]. Molecular ions are known to be cooled efficiently in sympathetic collisions with cold atoms or ions[6,7,8]. However, little knowledge is available on the elementary cooling steps, because the determination of quantum state-to-state collision rates at low temperature is prohibitively challenging for both experiment and theory. Here we present a method to manipulate molecular quantum states by non-resonant photodetachment. Based on this we provide absolute quantum scattering rate coefficients under full quantum state control for the rotationally inelastic collision of hydroxyl anions with helium. Experiment and quantum scattering theory show excellent agreement without adjustable parameters. Very similar rate coefficients are obtained for two different isotopes, which is linked to several quantum scattering resonances appearing at different energies. The presented method is also applicable to polyatomic systems and will help shed light on non-radiative processes in polyaromatic hydrocarbons and protein chromophores.**

* Email: roland.wester@uibk.ac.at

Molecular collisions at low temperatures have become known as a feature-rich quantum mechanical scattering problem. Theoretical calculations have shown already more than a decade ago that a plethora of scattering resonances become accessible once the thermal de Broglie wavelength of the collision system becomes comparable with the effective range of the intermolecular interaction[9]. Understanding the quantum dynamics of cold molecular interactions has created a rapidly growing field during the last decade[10,11,12]. Here, state-to-state inelastic collisions are among the most fundamental processes – and also the most difficult to study. Cold inelastic collisions of neutral OH molecules interacting with rare gas atoms[13] and, more recently, NO radicals[14] could be explored using the Stark decelerator technique. Cold chemical reactions of neutral molecules have been studied and the significance of isotope effects in quantum shape resonances[15] or the importance of tunnelling dynamics[16] have been identified. For molecular ions several cooling schemes have been developed for vibrational and rotational cooling**Fehler! Textmarke nicht definiert.**,[Fehler! Textmarke nicht definiert.,17,18,19]. Initial state-dependent collisions have previously been studied at higher temperature using laser-induced reactions[20,21], whereas rotational-state-resolved reactive collisions have recently been achieved at temperatures below 20 K**Fehler! Textmarke nicht definiert.**[,22].

Here, we present a versatile scheme to manipulate the rotational quantum states of negative ions – the state-selective removal of ions by threshold photodetachment - in order to study the fundamental inelastic process of rotational state-changing collisions. With this approach we determine the collision rate coefficients for the transition from the first excited J=1 state to the J=0 ground state of the hydroxyl negative ion and its deuterated isotopologue (see also Figure 1)

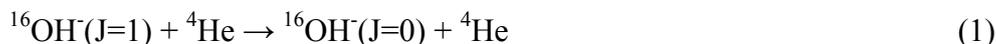  (1)

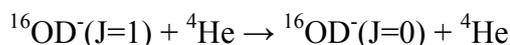

at a temperature of 15.3K[23]. The hydroxyl ions are perfect representations of a quantum mechanical rigid rotor, as neither electronic nor nuclear spin momentum couples appreciably to the mechanical rotation of the molecule. At long range the interaction potential does not distinguish between cations and anions, but at close proximity substantially different interaction energies are found as a function of distance and angle of approach. This influences reactive scattering of negative ions[24] and is also crucial for the studied inelastic collisions. For the OH⁻-He system it leads to a much weaker interaction at short range than for similar sized cations and suppresses the formation of bound OH⁻-He complexes even at 4K[25]. The interactions of the two isotopes OH⁻ and OD⁻ are described by the same anisotropic interaction potential[26], with the effect that different collision rates directly refer to differences in the nuclear quantum dynamics.

To selectively probe only the upper rotational state of the ion, an intense photodetachment laser beam depletes the excited J=1 state, but leaves the ground state anions intact. This laser drives the detachment reaction

$$^{16}OH^-(J=1) + h\nu \rightarrow {}^{16}OH(J=3/2) + e^- \quad (2)$$

$$^{16}OD^-(J=1) + h\nu \rightarrow {}^{16}OD(J=3/2) + e^-.$$

The effect of the two reactions (1) and (2) on an ensemble of cold trapped ions is described by a simple system of rate equations as depicted in Figure 1 and described in the supplementary information. In the weak intensity limit the detachment laser induces a loss rate for the molecular ions that is proportional to the ion fraction in the J=1 state and to the laser intensity[27,28]. For strong enough laser intensities the rate for photodetachment becomes comparable to the inelastic collision rate and induces a depletion of the excited state. In the experiment this leads to a deviation from the linear intensity dependence of the induced loss rate. Once the detachment rate is much larger than the inelastic collision rate, only the latter one determines the loss rate, since all molecular ions that reach the excited rotational state are immediately neutralised. In this discussion the effect of the second excited rotational state has been neglected, which is a safe assumption at the low temperatures in the ion trap (see Figure 1 and supplementary information). The degenerate magnetic sub-levels of the excited rotational states have not been treated separately, as they will mix rapidly by collisions and interactions with the trapping potential.

The measurements were performed on ions trapped inside a cryogenic multipole radiofrequency ion trap**Fehler! Textmarke nicht definiert.**. The experimental setup and the procedure of ion preparation and loading have already been presented earlier**Fehler! Textmarke nicht definiert.**[29], specific details are described in the supplementary information. Once loaded into the trap, a few hundred hydroxyl anions are being subjected to the photodetachment laser beam, the frequency of which is tuned 5-20cm⁻¹ below the electron affinity of OH (OD) (see Figure 1). This only neutralises OH⁻ (OD⁻) anions in J=1 (and above) and forms rotational ground state OH (OD) with total angular momentum J=3/2.

The exponential decay rates of trapped hydroxyl anions are determined from the stored ion signal as a function of the storage time in the trap for increasing intensity of the photodetachment laser. These rates are directly proportional to the population in the J=1 state. Plotting the ions' relative population as a function of their loss rate shows that with increasing loss rate, caused by increased photodetachment laser intensity, the excited state population is depleted. This is shown in Figure 2a and b for OH⁻ and OD⁻ for different helium densities in the ion trap. At the highest helium density the normalised population remains almost constant and thereby unaffected by the depletion action of the detachment laser, because the detachment rate remains weak compared to the inelastic collision rates. However, at lower helium densities a depletion of the excited J=1 state by more than 50% is observed. By fitting the rate equation model shown in Figure 1 to the data, the inelastic collision rates are extracted as a function of helium density. These are shown as a function of density in Figure 2c. The slope of the linear dependence yields the collision rate coefficients. They turn out to be almost the same for OH⁻ and OD⁻ and amount to $(6.4 \pm 1.5_{stat} \pm 2.1_{syst}) \times 10^{-11}$ cm³/s and $(4.8 \pm 1.6_{stat} \pm 1.6_{syst}) \times 10^{-11}$ cm³/s, respectively. The statistical accuracy of the two values is caused by the precision of the exponential decay rate measurements and the temperature determination by photodetachment spectroscopy, which is also traced back to decay rate measurements. The systematic accuracy stems from the absolute helium density calibration. The result for the ratio of the two rates yields $k_{OH^-}/k_{OD^-} = 1.3 \pm 0.5$, here the accuracy of the absolute density calibration cancels.

Both collision rate coefficients are smaller than the Langevin capture rate coefficient, which assumes unit scattering probability for all partial waves up to the maximum quantum number that allows the collision partners to overcome the centrifugal barrier. This Langevin rate is $5.8 \times 10^{-10}$ cm³/s for both OH⁻ and OD⁻ colliding with helium. Normalising the measured rates to the Langevin model, we find the inelastic collision probability for OH⁻ and OD⁻ to be about 10%. To understand why both isotopologues feature a similar, finite probability in inelastic collisions with helium, we have performed ab initio quantum scattering calculations for both anions on the same interaction potential surface**Fehler! Textmarke nicht definiert.** (neglecting non-Born Oppenheimer effects) using the time-independent quantum coupled-channel method[30]. In the framework of the rigid-rotor approximation the ground vibrational state of the hydroxyl anion and the first ten asymptotic rotational states were taken into consideration. Partial waves up to an angular momentum of 40ℏ were considered.

The calculations yield the collision energy-dependent integral scattering cross sections from which the thermal rate coefficients are obtained by proper averaging (shown in Figure 3a and b). The ratio of excitation and de-excitation rate coefficients are found to be in excellent agreement with the prediction of detailed balance. We also find that the two thermally averaged rate coefficients cross at about 17K. At lower temperature, OH$^-$ shows the larger rate coefficient, whereas at higher temperature the OD$^-$ rate becomes larger. This can be explained by the relative importance at different, but fairly close, collision energies of the numerous scattering resonances that appear in the energy-dependent cross section throughout the range of energies relevant at low temperature. The calculated rates at 15.3K (also plotted in Figure 3b) agree very well with the measured rate coefficients at this temperature. The ratio of the two calculated rate coefficients of 1.07 is found to be in perfect agreement with the experimental ratio. This is a manifestation of the precision that can be achieved in cold collisions when many scattering resonances have to be taken into account. It also confirms the limited predictive power of the simple Langevin model for ionic reactions at low temperatures.

Selective depletion of the excited rotational state of the hydroxyl anion is not only a means to measure inelastic relaxation rates and compare them with state-to-state ab-initio quantum cross section calculations. It can be used to lower the effective rotational temperature as a prerequisite for further quantum-state controlled experiments below the temperature of the neutral coolant. We have already achieved an effective rotational temperature for OD$^-$ of 10±1K for strong photodetachment laser powers and low buffer gas densities, which corresponds to a depletion of the excited J=1 state from 31% to 14% (see supplementary information). With further development excited state populations below 1% become feasible. More importantly, photodetachment depletion does not rely on specific molecular properties and is thereby directly applicable to complex polyatomic molecular ions, where it for example opens up a powerful approach to prepare internally cold organic molecules of astrophysical relevance, such as polyaromatic hydrocarbons, or of biophysical relevance, such as molecular chromophores of fluorescent proteins, for precision studies. Understanding inelastic collisions in these systems will provide new insight into the competition of collisional quenching with molecular line emission in radioastronomy and on non-radiative de-excitation of dye molecules.

**Acknowledgements**

This work has been supported by the European Research Council under ERC grant agreement No. 279898. E.E. acknowledges support from the Fond National de la Recherche Luxembourg.

**Author contributions**

D.H., S.L., O.L., S.S., E.E., S.K. performed the experiments. F.C. and F.G. carried out the calculations. R.W. planned and supervised the project. D.H., F.C., S.K., and R.W. prepared the manuscript with input from all authors.

**Additional information**

Supplementary information is available.

**Competing financial interests**

The authors declare no competing financial interests.

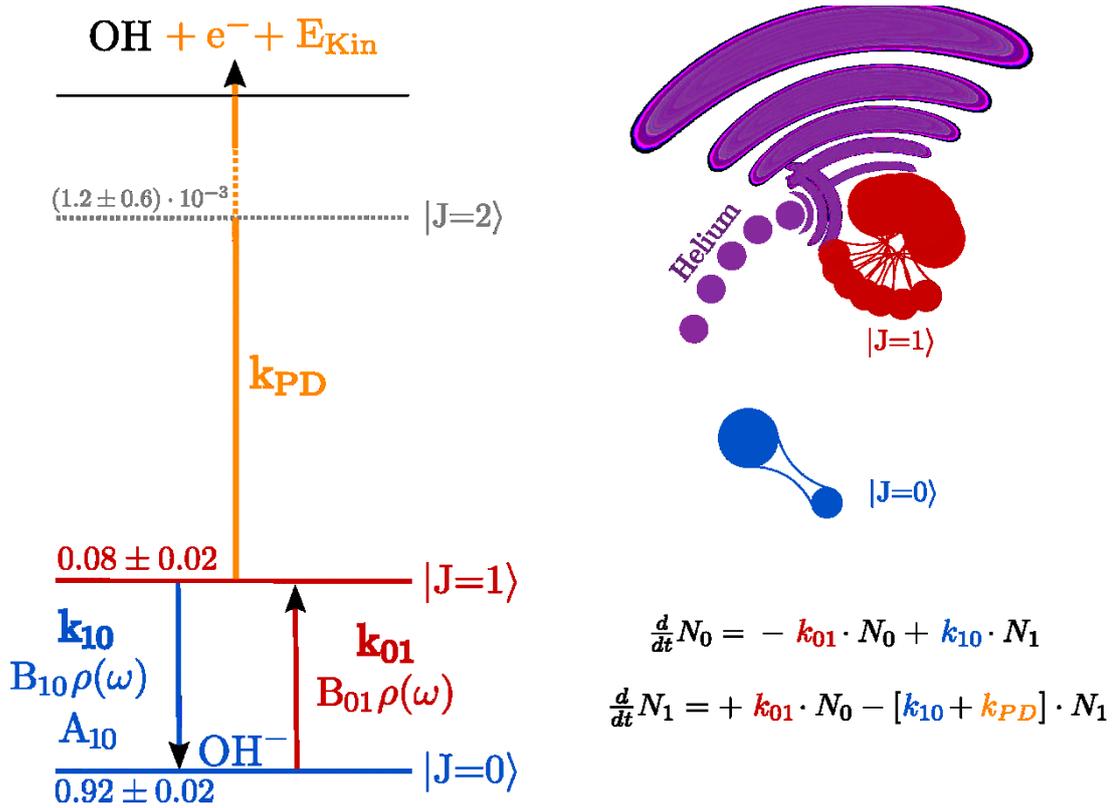

**Figure 1: OH⁻ rotational energy levels and coupling rates**

Diagram of the rotational energy levels of OH⁻, which are coupled by collisions with helium (with inelastic rate coefficients $k_{10}$ and $k_{01}$, see the artist's view on the right) and black-body radiation-induced transitions (with the Einstein coefficients $A_{10}$, $B_{10}$, and $B_{01}$). Photodetachment transitions with the rate $k_{PD}$ couple the excited anion states to the rotational ground state of neutral OH and a free electron. Given the population of the rotational levels, measured by photodetachment spectroscopy as described in the text, transitions involving J=2 can safely be neglected.

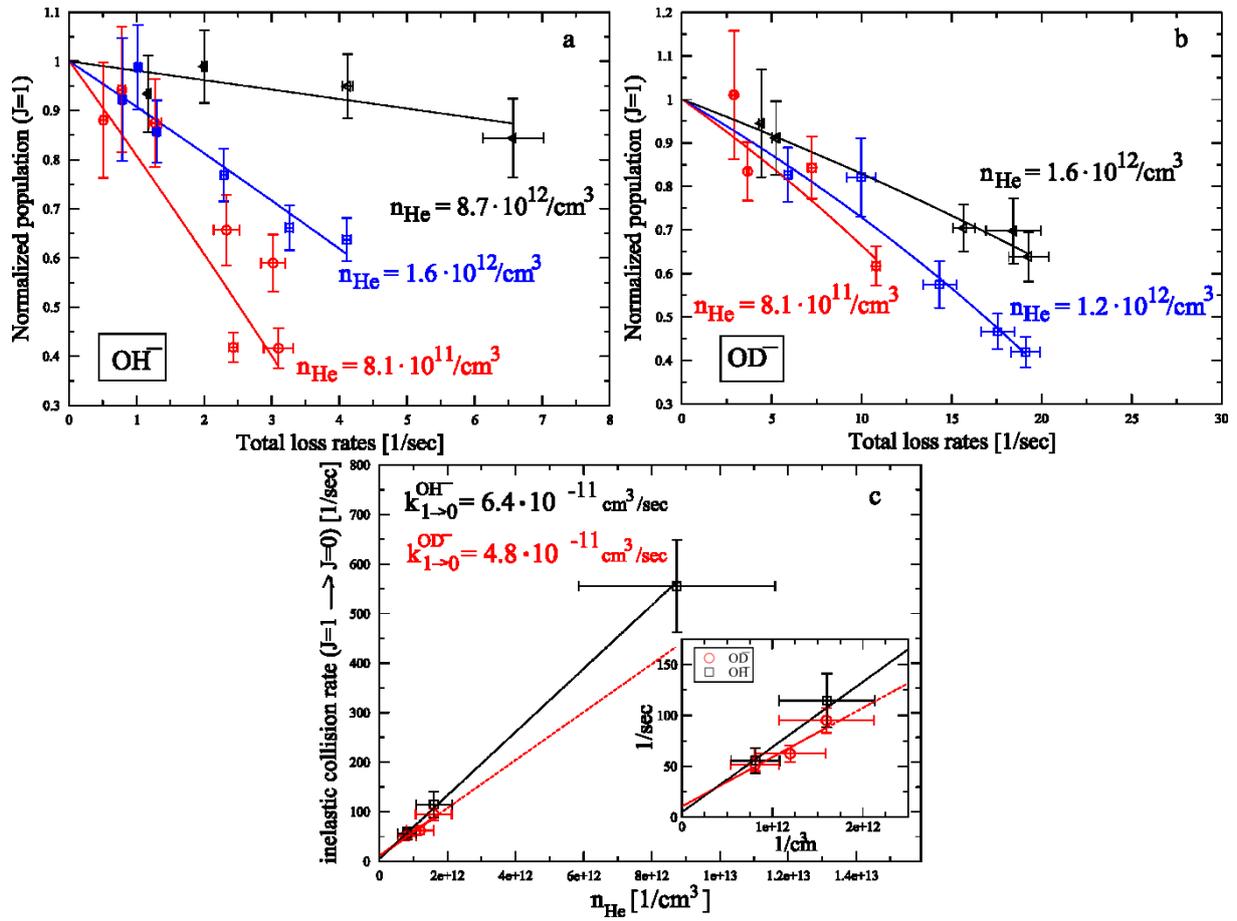

**Figure 2: Measured inelastic rate coefficients**

**a, b,** Plot of the relative populations of OH⁻ and OD⁻ ions in the J=1 rotational state versus the total loss rate induced by photodetachment for different helium buffer gas densities. The error bars are estimated standard deviation fit errors obtained in fitting exponential decay rates to the ion intensity as a function of storage time. The solid lines show fits to the solution of the rate equation system based on the parameters shown in Figure 1 with the inelastic rate $k_{10}$ as the only free parameter. **c,** Inelastic rates, as obtained from the fits to panels a and b as a function of the helium density. Vertical error bars are estimated fit errors; horizontal error bars show the estimated accuracy of the density measurements. The slopes of the linear fits to these data points represent the inelastic collision rate coefficients for OH⁻ and OD⁻. The inset shows a zoom of the data points at lower density.

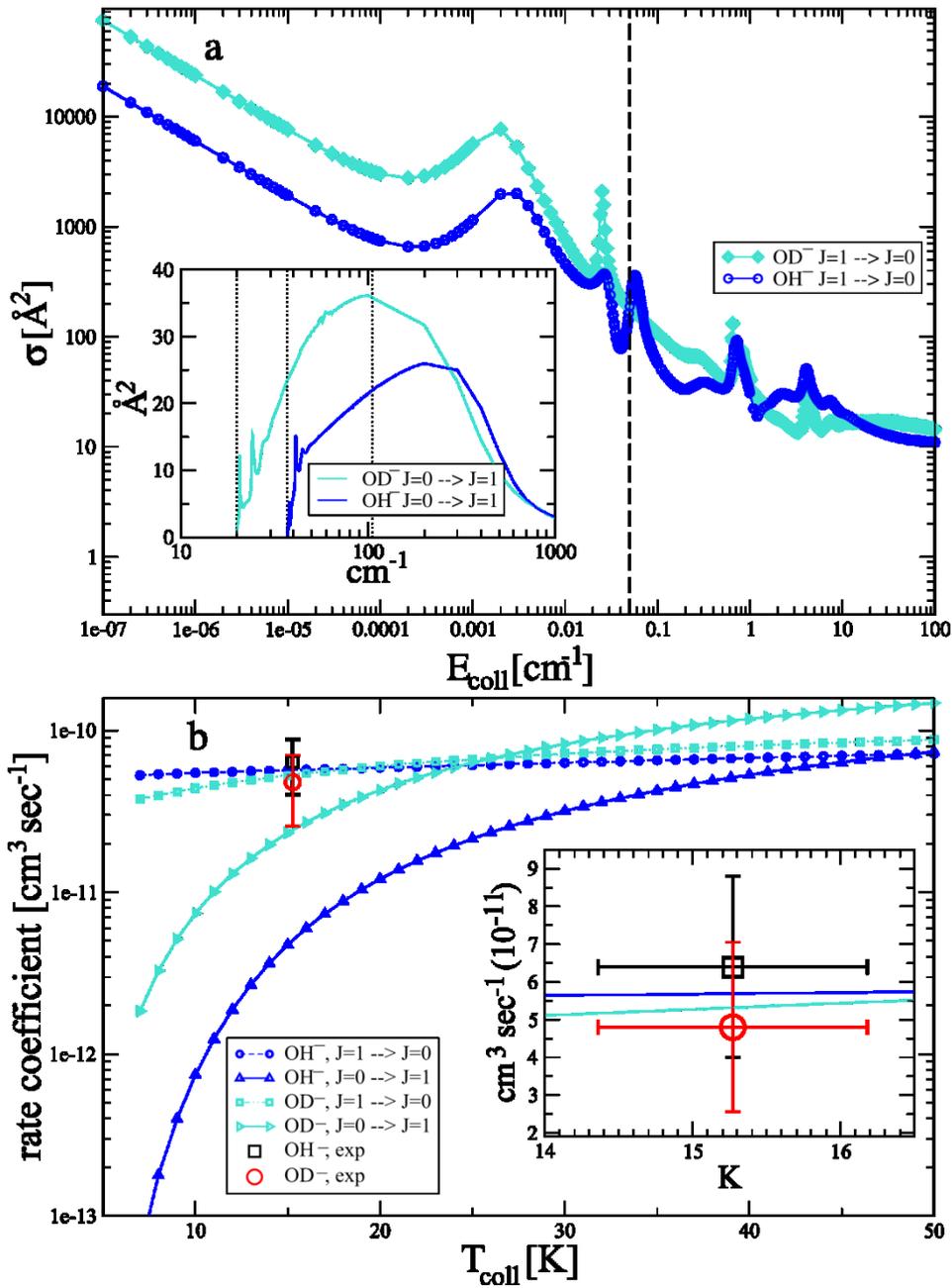

**Figure 3: Calculated inelastic cross sections and rate coefficients**

**a,** Inelastic scattering cross section for OH⁻ and OD⁻ transitions from J=1 to 0, calculated as a function of collision energy. Inset: Inelastic excitation cross sections from J=0 to 1. **b,** Temperature-averaged inelastic rate coefficients for excitation and de-excitation as a function of translational temperature. The dashed line in the upper panel marks the lowest relative energy that contributes significantly to the convolution of the rate thermal coefficients. Also shown are the measured rate coefficients with their respective error bars (see inset for a zoom with a linear axis scale).